\documentclass[aps,prl,twocolumn,showpacs,10pt,floatfix,longbibliography]{revtex4-2}

\usepackage{graphicx}
\usepackage{subfigure}
\usepackage{amsmath}
\usepackage{amsfonts}

\usepackage{mathrsfs}
\usepackage{bbm}
\usepackage{subfigure}
\usepackage{xcolor}
\usepackage[colorlinks=true, urlcolor=blue, linkcolor=blue, citecolor=blue]{hyperref}

\usepackage[T1]{fontenc}

\usepackage[per-mode=symbol]{siunitx}
\usepackage{bm}
\usepackage[version=4]{mhchem} 

\usepackage{stackengine}[2013-09-11]

\usepackage{dsfont}

\AddToHook{cmd/appendix/before}{%
  \setcounter{equation}{0}%
  \setcounter{theorem}{0}%
}

\begin{document}

\title{Dissipationless valley diode effects in Kekul\'e graphene superlattices}

\author{W.~Zeng}
\email{zeng@ujs.edu.cn}
\affiliation{Department of Physics, Jiangsu University, Zhenjiang 212013, China}

\begin{abstract}
Superconducting diodes have attracted considerable interest for nonreciprocal device applications owing to their dissipationless transport. Here, we theoretically propose that a tunnel junction based on the Kekul\'e graphene superlattice can host a finite valley supercurrent that exhibits pronounced nonreciprocity even in the absence of superconductivity, giving rise to a dissipationless valley diode effect. This effect originates from the additional reflection phase acquired by electrons in each transverse channel and supports perfect valley diode rectification with $100\%$ efficiency, which can be readily tuned via the gate voltage. This finding provides a platform for designing dissipationless valleytronics in a more versatile manner.
\end{abstract}
\maketitle

\textit{Introduction.--}
The superconducting diode effect refers to the nonreciprocal transport of supercurrents, in which the magnitudes of the critical currents differ for opposite current directions \cite{PhysRevLett.128.037001,nadeem2023superconducting,ando2020observation,PhysRevX.12.041013}. The \textit{charge} superconducting diode effect typically requires the breaking of both inversion symmetry and time-reversal symmetry (TRS) \cite{PhysRevResearch.4.033167,PhysRevLett.99.067004,PhysRevB.108.214519,fv8k-twwl,PhysRevApplied.18.034064,PhysRevLett.134.176002}, as demonstrated in various systems, including topological insulators \cite{PhysRevLett.131.096001,PhysRevB.106.214524,PhysRevB.109.075412,gc7k-rn8q} and altermagnets \cite{PhysRevB.110.024503,PhysRevB.110.014518,l42w-8qrz}. More recently, the \textit{spin} superconducting diode effect has extended this paradigm to spin transport. It can be realized in a Fulde-Ferrell superconducting chain \cite{PhysRevB.107.224510} or arise from the interplay between spin-orbit coupling and spin-triplet pairing \cite{PhysRevLett.132.216001}, resulting in direction-dependent critical spin supercurrents. In conventional superconducting diodes, dissipationless currents are intrinsically tied to the superconducting state, requiring the devices to operate below the superconducting transition temperature and thereby fundamentally limiting their operating temperature range. In addition, although TRS can be broken by intrinsic magnetism \cite{lin2022zero,PhysRevApplied.18.034064,golod2022demonstration}, switching the diode and reversing its polarity typically still require external magnetic fields \cite{jeon2022zero}, which can suppress superconductivity and induce vortex-motion dissipation \cite{PhysRevB.83.144526,Kopnin_2002}, compromising dissipationless transport. Therefore, achieving field-free diode effects with a wider operating temperature range remains a challenge \cite{qi2025high,ghosh2024high,10.1093/nsr/nwag285}.

Similar to the spin degree of freedom, the valley degree of freedom also offers a promising platform for information encoding and transport, forming the basis of valleytronics \cite{PhysRevLett.110.046601,PhysRevLett.116.016802,PhysRevLett.129.027401,PhysRevLett.117.257402,PhysRevLett.97.146805,PhysRevLett.127.196401,PhysRevLett.106.176802}. This naturally raises the question of whether analogous diode functionality can be realized in the valley degree of freedom, such that valley transport becomes direction dependent, enabling directional control of valley information.

\begin{figure}[tp]
\begin{center}
\includegraphics[clip = true, width =1\linewidth]{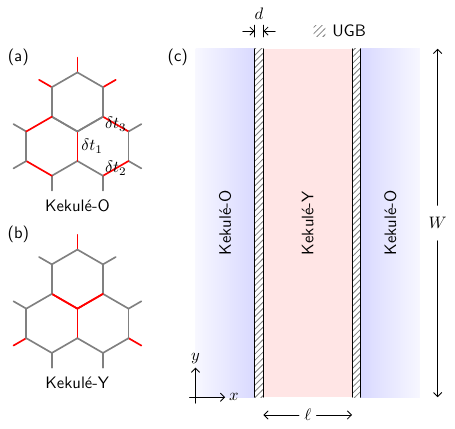}
\end{center}
\caption{[(a), (b)] Kekul\'e distortions in graphene. $\delta t_i$ ($i=1,2,3$) in (a) denotes three neighboring-bond modifications in Kekul\'e-O distortions. (c) Schematic of the tunnel junction. The blue and red regions represent the Kekul\'e-O and Kekul\'e-Y distorted graphene regions, respectively, with the latter having a length of $\ell$. The two shaded regions represent undistorted graphene barriers (UGBs), each of length $d$. }
\label{fig:jun}
\end{figure}

In this Letter, we report dissipationless nonreciprocal valley transport in a junction based on a Kekul\'e graphene superlattice (KGS) without relying on superconductivity. The Kekul\'e distortions in graphene can be realized by spatial modulation \cite{PhysRevLett.126.206804,ncq1-yzpd,gutierrez2016imaging,guan2024observation,PhysRevB.99.035411}. Specifically, the Kekul\'e-O distortion exhibits an O-shaped bond modulation pattern [Fig.~\ref{fig:jun}(a)] that opens a gap at the Dirac point, whereas the Kekul\'e-Y distortion features a Y-shaped pattern [Fig.~\ref{fig:jun}(b)] while preserving gapless Dirac cones with distinct Fermi velocities. By folding distinct valleys onto the center of the Brillouin zone, Kekul\'e distortions enable intriguing valley-dependent transport of low-energy carriers \cite{PhysRevB.98.195436,PhysRevB.105.125139,PhysRevB.104.075436,PhysRevB.105.094510,PhysRevB.106.165419,PhysRevB.97.241403,PhysRevB.101.245428}. We employ the Matsubara Green's function formalism to calculate the valley current in a KGS tunnel junction and find that the intervalley coupling induced by Kekul\'e distortions can give rise to a valley supercurrent with pronounced nonreciprocity, manifested by unequal critical currents in the forward and backward directions, thereby realizing a dissipationless valley diode effect. We further demonstrate that this effect originates from the additional phase acquired by back-reflected electrons in each transverse channel, and elucidate the role of symmetry in determining the emergence of this effect. By eliminating the need for superconductivity, the predicted effect circumvents the low superconducting transition temperature constraint inherent to conventional Josephson diodes. Furthermore, it operates without external fields and allows the valley diode efficiency to be readily controlled by the gate voltage, with perfect rectification reaching $100\%$ efficiency. These advantages open up a versatile avenue for realizing dissipationless valleytronic devices.

\textit{Model and formalism.--}
The KGS tunnel junction under consideration is shown in Fig.\ \ref{fig:jun}(c), where a Kekul\'e-Y distorted graphene region occupying $0<x<\ell$ is sandwiched between two  Kekul\'e-O distorted graphene regions. The low-energy excitations of the system are governed by the Hamiltonian $\hat{\mathcal{H}}_{\mathrm{KGS}}=\sum_{\bm k}\hat{c}^\dagger_{\bm k}\hat{\mathcal{H}}(\bm k)\hat{c}_{\bm k}$ with
\begin{align}
  \hat{\mathcal{H}}(\bm k)=\begin{pmatrix}
    \hbar v_F\bm k\cdot\hat{\bm{\sigma}}+U(x)&f_{\text{Kekul\'e}}(x)\\
    f^*_{\text{Kekul\'e}}(x)&-\hbar v_F\bm k\cdot\hat{\bm{\sigma}}+U(x)
  \end{pmatrix}.\label{ham}
\end{align}
where $\hat{c}_{\bm k}=(\hat{c}_{KA,\bm k},\hat{c}_{KB,\bm k},\hat{c}_{\bar{K}B,\bm k},\hat{c}_{\bar{K}A,\bm k})^T$ is the basis spinor containing both valley ($K,\bar{K}$) and sublattice pseudospin ($A,B$) degrees of freedom. The diagonal terms describe the pristine graphene, with $\bm k=(k_x,k_y)$ being the two-dimensional wave vector and $\hat{\sigma}_i$ ($i=x,y$) denoting Pauli matrices acting in pseudospin space. The electrostatic potential $U(x)=U\Theta(x)\Theta(\ell-x)$ is applied to the central region to tune its local potential energy, which can be adjusted by a gate voltage. The off-diagonal element $f_{\text{Kekul\'e}}(x)$ captures the effect of the Kekul\'e distortions in the junction.

The Kekul\'e-O distortion in the junction introduces a complex intervalley coupling $\Delta e^{i\phi}$ \cite{gamayun2018valley,PhysRevB.97.241403}, where $\Delta$ and $\phi$ are related to the hopping modifications by $\Delta\cos\phi=\delta t_1-\delta t_2/2-\delta t_3/2$ and $\Delta\sin\phi=\sqrt{3}(\delta t_2-\delta t_3)/2$ \cite{PTPS176253,PhysRevB.89.245421}. Here, $\delta t_{1,2,3}$ denote the nearest-neighbor hopping modifications, as shown in Fig.\ \ref{fig:jun}(a). This intervalley coupling gives rise to a $\bm{k}$-independent term in Eq.\ (\ref{ham}), given by $f_{\text{Kekul\'e}}(x)=\Delta(\cos\phi_j\hat{\tau}_x-\sin\phi_j\hat{\tau}_y)$ for $x<-d$ and $x>\ell+d$, where $\hat{\tau}_{x,y}$ are Pauli matrices acting in valley space. We assume equal magnitudes but different phases of the intervalley coupling in the left (L) and right (R) Kekul\'e-O distorted regions, denoted by $\phi_j$ ($j=L,R$). This distortion opens a gap of $2\Delta$, with doubly degenerate bands described by $E=\alpha\sqrt{(\hbar v_F |\bm k|)^2+\Delta^2}$, where $\alpha=\pm1$ labels the conduction and valence bands, respectively. In contrast to the superconducting gap, which is typically on the order of a few $\unit{\milli\electronvolt}$, the band gap in Kekul\'e-O graphene due to the intervalley coupling can reach several hundred $\unit{\milli\electronvolt}$ \cite{PhysRevLett.131.266501,gamayun2018valley}. For example, a gap of $2\Delta \approx \SI{200}{\milli\electronvolt}$ has been experimentally observed in graphene with Kekul\'e-O order induced by self-assembled Li adatoms \cite{sciadvabm5180}.

The Kekul\'e-Y distortion in the central region induces a valley-momentum coupling of the form \cite{gamayun2018valley} $f_{\text{Kekul\'e}}(x)=-\zeta\hbar v_F(k_x\hat{\tau}_x+k_y\hat{\tau}_y)\hat{\sigma}_z$ for $0<x<\ell$, where $0<\zeta<1$ denotes the coupling strength. The energy dispersion remains gapless, with $E_\pm=\alpha\hbar v_F(1\pm\zeta)|\bm k|$. Additionally, two undistorted graphene barriers (UGBs), each consisting of a pristine graphene region of length $d$ subject to a potential barrier of height $V$, are introduced at the Kekul\'e-Y/O interfaces [shaded regions in Fig.\ \ref{fig:jun}(c)]. Since the intervalley coupling is absent in these regions, the valley currents can be readily evaluated therein.

We assume the translation symmetry along the $y$ direction, so that $k_y=2n\pi/W$ is a good quantum number with $W$ the junction width and $n$ an integer. Thus, the Hamiltonian of the KGS junction can be mapped onto a set of $k_y$-dependent (quasi-)one-dimensional (1D) chains with equal spacing $a$, such that $\hat{\mathcal{H}}_{\mathrm{KGS}}=\sum_{k_y}\hat{\mathcal{H}}_{\mathrm{1D}}(k_y)$, where
\begin{align}
  \hat{\mathcal{H}}_{\mathrm{1D}}=\hat{\mathcal{H}}_0+\hat{\mathcal{H}}_L+\hat{\mathcal{H}}_R+\hat{\mathcal{H}}_C+\hat{\mathcal{H}}_{V}.\label{eq:1d}
\end{align}
Here $\hat{\mathcal{H}}_0=\sum_l[\hat{c}^\dagger_lt\sin k_ya\hat{\tau}_z\hat{\sigma}_yc_l+\hat{c}^\dagger_l(t\hat{\tau}_z\hat{\sigma}_x/2i)\hat{c}_{l+1}+h.c.]$ accounts for the pristine graphene substrate across the entire system with $\hat{c}_l=(\hat{c}_{K,l},\hat{c}_{\bar{K},l})^T$ and the hopping energy is given by $t=\hbar v_F/a$. $\mathcal{H}_{j}$ ($j=L,R$) represent the additional terms arising from the Kekul\'e-O distortions in the left (L) and right (R) leads, which are given by $\hat{\mathcal{H}}_{j}=\sum_{l\in j}\hat{c}^\dagger_l(\Delta\cos\phi_{j}\hat{\tau}_x-\Delta\sin\phi_{j}\hat{\tau}_y)\hat{c}_l$. We take $\phi_L=0$ and $\phi_R=\phi$ in our model. The Kekul\'e-Y distortions in the central region gives rise to $\hat{\mathcal{H}}_{C}=\sum_{l\in C}[\hat{c}^\dagger_l(t\zeta\sin k_ya\hat{\tau}_y\hat{\sigma}_z+U)\hat{c}_l+\hat{c}^\dagger_l(it\zeta\hat{\tau}_x\hat{\sigma}_z/2)\hat{c}_{l+1}+h.c.]$. For the two UGBs, each having a length of $\delta_x a$ ($\delta_x a=d$), we consider a potential barrier of height $V$, described by $\hat{\mathcal{H}}_V=\sum_{l}V\hat{c}^\dagger_l \hat{c}_l$. For small $\delta_x$, $V\delta_x/t$ is proportional to the strength of the Dirac delta-function potential localized at the interface in the continuum model \cite{PhysRevB.110.085301,PhysRevB.105.054502,PhysRevLett.97.217001}. We choose $\delta_x=2$ in our model, and this choice does not affect the main conclusions of our study.

The valley supercurrent flowing through the junction is given by $J_\mathrm{v}(\phi)=\sum_{k_y}j_\mathrm{v}(\phi)$, with $j_\mathrm{v}(\phi)$ denoting supercurrent of the $k_y$-dependent 1D junction described by $\hat{\mathcal{H}}_{\mathrm{1D}}$. This quantity can be calculated in terms of the Matsubara Green's function [see Sec.\ S1 in Supplement Material (SM) \cite{SM}]
\begin{align}
    j_{\mathrm{v}}(\phi)=\frac{4e}{\beta\hbar}\sum_{\omega_n}\operatorname{Im}\{\operatorname{Tr}g_{K\bar{K}}\hat{t}^\dagger_{\bar{K}}\mathcal{G}_{\bar{K}K}\hat{t}_K\},
\end{align}
where $\beta=1/k_BT$ is the inverse temperature, $\omega_n=(2n+1)\pi/\beta$ is the Matsubara frequency, $g_{K\bar{K}}$ denotes the anomalous Green's function at site $l$ with all sites to its right removed, whereas $\mathcal{G}_{\bar{K}K}$ is the full anomalous Green's function at site $l$, $\hat{t}^\dagger_{\tau}$ ($\tau=K,\bar{K}$) is the $\tau$-component of the coupling matrix between the ($l$)th and ($l+1$)th sites, which can be read off directly from Eq.\ (\ref{eq:1d}).

We first examine the $k_y$-resolved valley supercurrent $j_\mathrm{v}(\phi)$ in the 1D junction described by $\hat{\mathcal{H}}_{\mathrm{1D}}$, then elucidate the physical origin of its nonreciprocity, and finally sum over all $k_y$ modes to obtain the total current, thereby demonstrating the emergence of an gate-tuned dissipationless valley diode effect. Throughout the numerical calculations, we set $a=\Delta=1$ as the length and energy units, respectively.

\begin{figure}[tp]
\begin{center}
\includegraphics[clip = true, width =\linewidth]{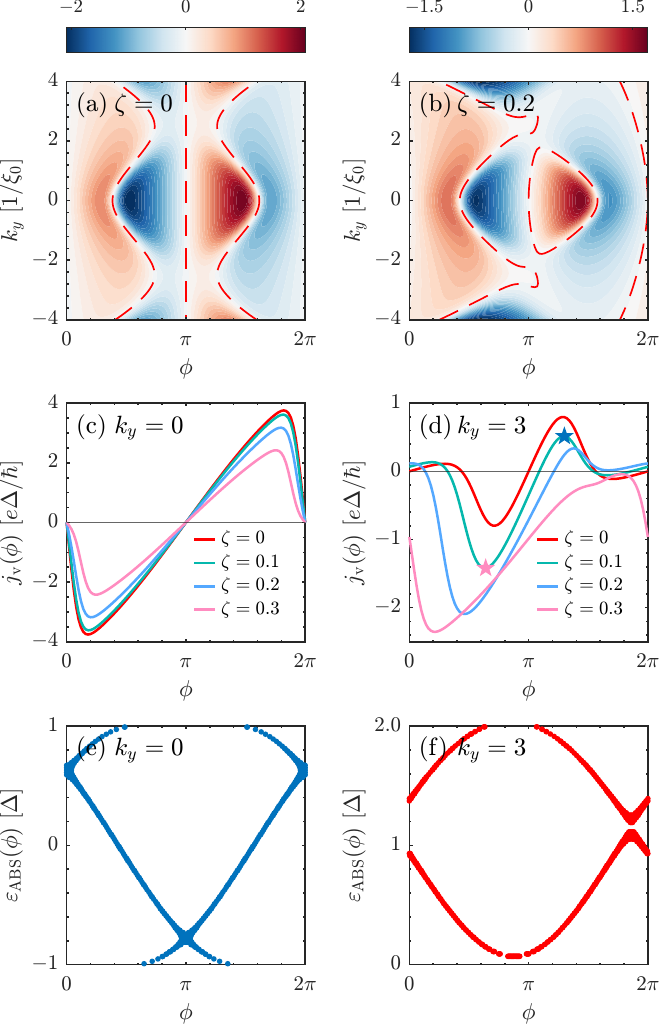}
\end{center}
\caption{[(a), (b)] Color map of the valley supercurrent $j_{\mathrm{v}}$ as a function of $k_y$ and $\phi$ at $V/t=0.25$. The color intensity represents the absolute value of the current and the red dashed line marks the zero-current point ($j_{\mathrm{v}}=0$). [(c), (d)] CPR of $j_{\mathrm{v}}(\phi)$ with different values of $\zeta$ at $V/t=0.11$. The blue and pink stars in (d) indicate the forward and backward critical currents for $\zeta=0.1$, respectively. [(e), (f)] Andreev-like levels $\varepsilon_{\mathrm{ABS}}$ as a function of $\phi$ at $\zeta=0.2$. The parameters are set as $t/\Delta=200$, $U/\Delta=10$, $\ell=24a$, and $T=0.05\Delta/k_B$. $k_y$, $j_{\mathrm{v}}$, and $\varepsilon_{\mathrm{ABS}}$ are given in units of $1/\xi_0$, $e\Delta/\hbar$, and $\Delta$, respectively, where the coherence length is estimated to be $\xi_0=\hbar v_F/\Delta=200a$.}
\label{fig:jv}
\end{figure}

\textit{Valley supercurrent in the $k_y$-dependent 1D junction.--}
Figs. \ref{fig:jv}(a) and \ref{fig:jv}(b) present $j_{\mathrm{v}}$ as a function of the transverse wave number $k_y$ and the phase $\phi$. In the absence of the Kekul\'e-Y distortion in the central region ($\zeta=0$), $j_\mathrm{v}$ is an even function of $k_y$. For each given $k_y$, the magnitudes of the currents in opposite directions are symmetric about $\phi=\pi$, indicating reciprocal transport; see Fig.\ \ref{fig:jv}(a). When the Kekul\'e-Y distortion is present in the central region, $j_\mathrm{v}$ remains even in $k_y$, whereas the magnitude is no longer symmetric about $\phi=\pi$, as shown in Fig.\ \ref{fig:jv}(b) for $\zeta=0.2$. This asymmetry indicates unequal critical currents in opposite directions, implying nonreciprocal transport.

To further illustrate the nonreciprocity, we present the current-phase relations (CPRs) of the valley supercurrent in Figs.\ \ref{fig:jv}(c) and \ref{fig:jv}(d). Fig.\ \ref{fig:jv}(c) presents the CPR at normal incidence ($k_y=0$) with several different $\zeta$. In the absence of Kekul\'e-Y distortion ($\zeta=0$), the CPR exhibits a sinusoidal-like dependence with equal forward and backward critical currents, where $j_\mathrm{v}(\phi)$ vanishes identically at $\phi=0$ and $\pi$, as shown by the red curve. Upon introducing the Kekul\'e-Y distortion, the overall magnitude of the CPR is modified, while the forward and backward critical currents remain equal in magnitude, and thus no nonreciprocal response arises.

Fig. \ref{fig:jv}(d) shows the CPR for oblique-incidence modes (e.g., $k_y=3$). For vanishing Kekul\'e-Y distortion, the CPR exhibits symmetric behavior similar to that at $k_y=0$, with equal magnitudes of the forward and backward critical currents (red curve). However, a finite $\zeta$ renders the CPR asymmetric. At $\zeta=0.1$ (green curve), the forward critical current is approximately $0.5$ (marked by the blue star), whereas the magnitude of the backward critical current exceeds $1$ (marked by the pink star), indicating nonreciprocal behavior between the two directions. The nonreciprocity becomes more pronounced as $\zeta$ increases, with the forward critical current decreasing and the backward critical current increasing in magnitude. At $\zeta=0.3$, the forward current is completely suppressed, while only the backward current is allowed to flow, resulting in perfect rectification (pink curve).

The nonreciprocity induced by the Kekul\'e-Y distortion is also manifested in the Andreev-like bound state (ABS). Figs.\ \ref{fig:jv}(e) and \ref{fig:jv}(f) show the $\phi$-dependent ABS spectrum $\varepsilon_{\mathrm{ABS}}(\phi)$ at $\zeta=0.2$, obtained either by the transfer matrix formalism, or by searching the peaks of the local density of states; see Sec.\ S2 in SM \cite{SM}. For $k_y=0$, $\varepsilon_{\mathrm{ABS}}(\phi)$ is symmetric under $\phi\to 2\pi-\phi$ [Fig.\ \ref{fig:jv}(e)]. In contrast, this symmetry is broken for oblique incidence, as illustrated in Fig.\ \ref{fig:jv}(f) for $k_y=3$, indicating an asymmetric CPR. Owing to the absence of particle-hole symmetry, $\varepsilon_{\mathrm{ABS}}(\phi)$ lacks the symmetry about $E=0$ that characterizes a conventional Josephson junction.

\textit{Additional reflection phase induced nonreciprocity.--}
The nonreciprocal behavior of the $k_y$-resolved $j_{\mathrm{v}}(\phi)$ can be understood in terms of the additional reflection phase acquired by the electrons in each $k_y$ channel. Analogous to the geometric phase in optics \cite{Berry01111987,PhysRevLett.126.183902,PhysRevA.109.023517}, an electron can acquire an additional phase in its reflection amplitude when scattered at a heterojunction interface, owing to the rotation of its wavefunction in the space spanned by its internal degrees of freedom. This additional phase is governed by pseudospin-momentum locking in undistorted graphene tunnel junctions and depends on the transverse momentum $k_y$, giving rise to various intriguing transverse transport phenomena \cite{physRevB.87.165420,PhysRevB.110.205406}. However, in our model, the valley-momentum locking induced by Kekul\'e-Y distortion makes the reflection phase dependent on the intervalley coupling phase $\phi$, thereby modifying the valley supercurrent conjugate to $\phi$. We show below that this acquired phase is the key ingredient underlying nonreciprocal valley transport.

\begin{figure}[tp]
\begin{center}
\includegraphics[clip = true, width =\linewidth]{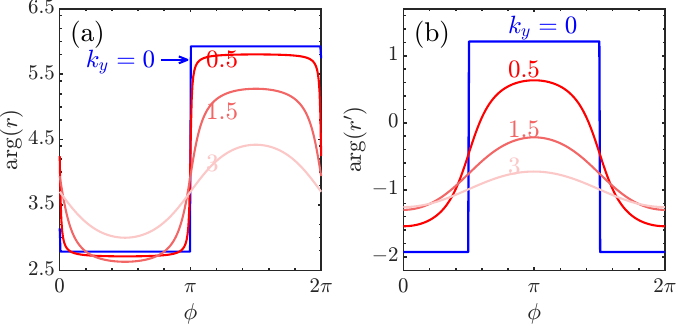}
\end{center}
\caption{The reflection phase for (a) intraband reflections and (b) interband reflections with $U/\Delta=10$ and $V/t=0.25$. Curves in different colors correspond to different $k_y$ values, labeled alongside each curve in units of $1/\xi_0$.  }
\label{fig:phase}
\end{figure}

The $\phi$-dependent reflection phase can be obtained by matching the wave functions at the junction interface; see Sec.\ S2 in SM \cite{SM}. The intraband reflection phase ($\arg r$) and interband reflection phase ($\arg r'$) as functions of $\phi$ for different $k_y$ are shown in Figs.\ \ref{fig:phase}(a) and \ref{fig:phase}(b), respectively. At $k_y=0$, both intraband and interband reflection phases are independent of $\phi$, up to isolated $\pi$-jumps; see blue curves in Fig.\ \ref{fig:phase}. In contrast, for $k_y\neq 0$, the reflection phases depend nonlinearly on $\phi$ (red curves). The Andreev-like levels for the $k_y$-dependent 1D junction can be qualitatively written as $\varepsilon_{\pm}(\phi)=\pm \varepsilon_{s}(\phi)+\varepsilon_0$, where $\varepsilon_0$ is a $\phi$-independent energy shift due to the broken of the particle-hole symmetry and~\cite{PhysRevB.74.041401}
\begin{align}
  \varepsilon_s(\phi)=\Delta\sqrt{1-t_s\sin^2\theta_s(\phi)}.\label{eq:bs}
\end{align}
Here $t_s$ ($s=\pm$) are the transmissions at $\Delta=0$ for the $k_y$ mode and
\begin{align}
\theta_s(\phi)=[\phi+\phi_s+s\Gamma(\phi)]/2,\label{eq:gxc1}
\end{align}
where $\phi_s=2(k_s^>-k_{-s}^<)\ell$ is the kinetic phase due to the motion of electrons in the central region with $k_s^>$ ($k_s^<$) being the positive (negative) propagating longitudinal wave vector for $E_s$ band in the Kekul\'e-Y region. $\Gamma(\phi)$ is the total additional reflection phase acquired at the two interfaces of the junction. As bound-state formation involves both intraband and interband reflection, the corresponding reflection phases indicate that $\Gamma$ is also $\phi$-independent at $k_y=0$, but depends nonlinearly on $\phi$ for $k_y\neq0$. This nonlinear dependence on $\phi$ can give rise to nonreciprocity in the valley supercurrent conjugate to $\phi$, which is given by \cite{brouwer1997anomalous}
\begin{align}
  j_{\mathrm{v}}(\phi)=-(2e/\hbar)\sum_s\partial_\phi\varepsilon_s(\phi)\tanh(\beta\varepsilon_s/2).\label{eq:gxc2}
\end{align}

To make the role of $\Gamma(\phi)$ in the nonreciprocal transport more explicit, we qualitatively parametrize it as
\begin{align}
  \Gamma(\phi)=w(\phi)\sin(\phi/2)+w_0.\label{eq:gxc3}
\end{align}
Here $w(\phi)$ is a $\phi$-dependent parameter that vanishes at $k_y=0$ and varies with $k_y$ for $k_y\neq 0$, while $w_0$ is a $\phi$-independent parameter. When $w(\phi)$ is nonsingular and slowly varying over $[0,2\pi)$, $\Gamma(\phi)$ can be approximated as the additional phase arising from interband reflection, which attains extrema at $\phi=0,\pi$; see Fig.\ \ref{fig:phase}(b). Eq.\ (\ref{eq:bs}) then provides a qualitative description of the Andreev-like levels dominated by the interband process. Combining Eqs.\ (\ref{eq:bs}-\ref{eq:gxc3}), we obtain the valley supercurrent for $k_y$ mode
\begin{align}
  j_{\mathrm{v}}(\phi)&=\frac{e\Delta}{2\hbar}\sum_{s=\pm}\frac{t_s\sin2\theta_s}{\sqrt{1-t_s\sin^2\theta_s}}\big(1-s w\cos(\phi/2)\big),
\end{align}
which directly yields the difference between the forward critical current ($j_{\mathrm{v}}^{c+}$) and backward critical current ($j_{\mathrm{v}}^{c-}$)
\begin{align}
   \delta j_{\mathrm{v}}^c=j_{\mathrm{v}}^{c+}-|j_{\mathrm{v}}^{c-}|\propto w\sin(\delta k_F\ell),\label{eq:qjc} 
\end{align}
with $\delta k_F=2U/\hbar v_F$; see Sec.\ S3 in SM \cite{SM}. Nonreciprocity thus emerges for $w\neq 0$, corresponding to $k_y\neq 0$, which explains our numerical results in Figs.\ \ref{fig:jv}(c-d).

\textit{Symmetry analysis.--}
We now examine the symmetry constraints on the emergence of the nonreciprocal valley supercurrent. Since Kekul\'e distortions preserve TRS, the Hamiltonian satisfies
\begin{align}
  \hat{\mathcal{T}}\hat{\mathcal{H}}(\bm k,\phi)\hat{\mathcal{T}}^{-1}=\hat{\mathcal{H}}(-\bm k,\phi),\label{eq:TRS}
\end{align}
which, in the $k_y$-dependent 1D representation, becomes $\hat{\mathcal{T}}\hat{\mathcal{H}}_{\mathrm{1D}}(k_y,\phi)\hat{\mathcal{T}}^{-1}=\hat{\mathcal{H}}_{\mathrm{1D}}(-k_y,\phi)$. Here $\hat{\mathcal{T}}=\hat{\tau}_x\hat{\sigma}_x\hat{\mathcal{K}}$ is the TRS operator with $\hat{\mathcal{K}}$ being the complex conjugation. In the presence of TRS, the Andreev-like levels obtained from $\hat{\mathcal{H}}_{\mathrm{1D}}(k_y,\phi)\psi_s=\varepsilon_s(k_y,\phi)\psi_s$ imply that $\hat{\mathcal{T}}\psi_s$ satisfies $\hat{\mathcal{H}}_{\mathrm{1D}}(-k_y,\phi)\hat{\mathcal{T}}\psi_s=\varepsilon(k_y,\phi)\hat{\mathcal{T}}\psi_s$. Consequently, $\varepsilon_s(k_y,\phi)$ is an even function of $k_y$, indicating that $j_{\mathrm{v}}$ [proportional to $\sum_s\partial_\phi\varepsilon_s(\phi)$] is also even in $k_y$, which is consistent with the numerical results shown in Figs.\ \ref{fig:jv}(a) and \ref{fig:jv}(b). This scenario is fundamentally differs from that in conventional Josephson junctions, where TRS reverses the superconducting phase difference $\phi_{\mathrm{sc}}$, yielding the asymmetric current $j(\phi_{\mathrm{sc}})=-j(-\phi_{\mathrm{sc}})$, and thus TRS breaking is generally required for a nonreciprocal Josephson effect. In contrast, since TRS leaves $\phi$ invariant in our model, nonreciprocal valley supercurrents can arise without TRS breaking. Furthermore, we can show that the net charge current in the system always vanishes at equilibrium, irrespective of the presence or absence of TRS; see Sec.\ S1 in SM \cite{SM}. This indicates that external-field perturbations do not generate any spurious charge current contributions, allowing the system to robustly sustain pure valley current transport.

In fact, in the absence of the Kekul\'e-Y distortion, the Hamiltonian preserves a combined valley-sublattice symmetry $\hat{\mathcal{U}}=\hat{\tau}_x\hat{\sigma}_z$ under the reversal of $\phi$
\begin{align}
    \hat{\mathcal{U}}\hat{\mathcal{H}}(\bm k,\phi)\hat{\mathcal{U}}^{-1}=\hat{\mathcal{H}}(\bm k,-\phi).
\end{align}
As discussed above, this $\hat{\mathcal{U}}$ symmetry gives rise to $j_{\mathrm{v}}(k_y,\phi)=-j_{\mathrm{v}}(k_y,-\phi)$, and thus prohibits the emergence of the nonreciprocal valley supercurrent. However, the Kekul\'e-Y distortion breaks $\hat{\mathcal{U}}$ symmetry with the nonezero $k_y$ term $\zeta k_y\hat{\tau}_y\hat{\sigma}_z$. Thus, in the presence of Kekul\'e-Y distortion, nonreciprocal valley currents can emerge only for $k_y\neq 0$. The symmetry analysis is consistent with our numerical results and the above analysis of the additional reflection phase.

\begin{figure}[tp]
\begin{center}
\includegraphics[clip = true, width =\linewidth]{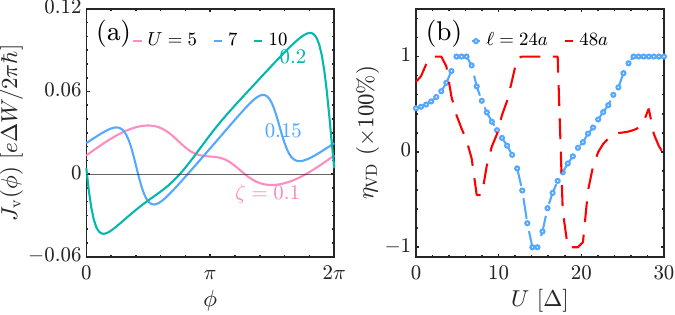}
\end{center}
\caption{(a) CPR of the net valley supercurrent $J_{\mathrm{v}}$ with $V/t=0.11t$ and $\ell=24a$. (b) The valley diode efficiency $\eta_{\mathrm{VD}}$ as a function of $U$ with $\zeta=0.1$ and $V/t=0.11$.   }
\label{fig:qv}
\end{figure}

\textit{Gate-tuned valley diode effects.--}
The net valley supercurrent $J_\mathrm{v}(\phi)$ is obtained by summing $j_\mathrm{v}(\phi)$ over all $k_y$ modes. As shown in Fig.\ \ref{fig:qv}(a), the resulting CPR of $J_\mathrm{v}(\phi)$ retains the nonreciprocity of the individual $j_\mathrm{v}(\phi)$ contributions at finite $k_y$, yielding unequal forward and backward critical currents, $J_\mathrm{v}^{c+}\neq |J_\mathrm{v}^{c-}|$. For $\zeta=0.1$, $|J_\mathrm{v}^{c+}|$ is more than three times larger than $J_\mathrm{v}^{c-}$ (pink curve). The asymmetry between $J_\mathrm{v}^{c+}$ and $J_\mathrm{v}^{c-}$ persists as $U$ and $\zeta$ vary. In fact, the extent of the nonreciprocity can be tuned by varying the gate voltage $U$, as quantified by the valley diode efficiency
\begin{align}
  \eta_{\mathrm{VD}}=\frac{J_\mathrm{v}^{c+}-|J_\mathrm{v}^{c-}|}{J_\mathrm{v}^{c+}+|J_\mathrm{v}^{c-}|}.
\end{align}
Fig. \ref{fig:qv}(b) shows the dependence of $\eta_{\mathrm{VD}}$ on the gate voltage $U$. As $U$ increases, $\eta_{\mathrm{VD}}$ exhibits periodic behaviors reminiscent of the Fabry-P\'erot oscillations \cite{PhysRevLett.101.156804,campos2012quantum}, which imply the essential role of the additional reflection phase in the emergence of the nonreciprocity. As revealed by our qualitative analysis of the ABSs, the oscillations originate from the $\sin(\delta k_F\ell)$ factor, which provides a qualitative estimate of the oscillation period, $\pi at/\Delta\ell\approx13$ for $\ell=48a$ (red). This period doubles as $\ell$ decreases to $24a$ (blue). By tuning $U$, the perfect valley diode rectification with $\eta_{\mathrm{VD}}=100\%$ can be achieved.

\textit{Summary.--}
In conclusion, we theoretically propose a dissipationless valley diode effect in a tunnel junction based on Kekul\'e graphene superlattices. Its nonreciprocity originates from the additional reflection phase accumulated by electrons in each transverse channel and does not require the breaking of TRS. In contrast to conventional superconducting diode effects, the predicted effect does not rely on superconductivity or external fields, and its valley diode efficiency can be readily tuned by adjusting the gate voltage, with perfect rectification achieving an efficiency of up to $100\%$, thereby providing greater flexibility in device design and highlighting its potential for future applications in valleytronics.

\begin{acknowledgments}
This work was supported by the National Natural Science Foundation of China (Grant No.\ 12504052), the Natural Science Foundation of Jiangsu Province (Grant No.\ BK20250838), and the Natural Science Foundation of the Jiangsu Higher Education Institutions of China (Grant No.\ 25KJB140004).
\end{acknowledgments}

\end{document}


\title{Supplementary Material for ``Dissipationless valley diode effects in Kekul\'e graphene superlattices''}

\author{W.~Zeng}
\email{zeng@ujs.edu.cn}
\affiliation{Department of Physics, Jiangsu University, Zhenjiang 212013, China}

\maketitle

\setcounter{section}{0}
\tableofcontents

\section{Derivation of the valley supercurrent using the Matsubara Green's Function}\label{s1}
In this section, we derive an expression for the valley supercurrent in the KGS junction using the Matsubara Green's function formalism. The junction Hamiltonian $\hat{\mathcal{H}}_{\mathrm{KGS}}$ can be mapped onto a set of $k_y$-dependent 1D chains with equal spacing $a$, such that $\hat{\mathcal{H}}_{\mathrm{KGS}}=\sum_{k_y}\hat{\mathcal{H}}_{\mathrm{1D}}(k_y)$, where 
\begin{align}
  \hat{\mathcal{H}}_{\mathrm{1D}}(k_y)&=\sum_l[\hat{c}^\dagger_lt\sin k_ya\hat{\tau}_z\hat{\sigma}_yc_l+\hat{c}^\dagger_l(t\hat{\tau}_z\hat{\sigma}_x/2i)\hat{c}_{l+1}]+\sum_{l\in L/R}\hat{c}^\dagger_l(\Delta\cos\phi_{L/R}\hat{\tau}_x-\Delta\sin\phi_{L/R}\hat{\tau}_y)\hat{c}_l\nonumber\\
  &\quad+\sum_{l\in C}[\hat{c}^\dagger_l(t\zeta\sin k_ya\hat{\tau}_y\hat{\sigma}_z+U)\hat{c}_l+\hat{c}^\dagger_l(it\zeta\hat{\tau}_x\hat{\sigma}_z/2)\hat{c}_{l+1}]+\sum_{l\in\mathrm{UGBs}}V\hat{c}^\dagger_l \hat{c}_l+h.c..\label{eq:h1d}
\end{align}
Here, the hopping energy is given by $t=\hbar v_F/a$. The operator spinor $\hat{c}_l$, which incorporates both valley and sublattice degrees of freedom, is defined as $\hat{c}_l=(\hat{c}_{K,l},\hat{c}_{\bar{K},l})^T$ with $\hat{c}_{K,l}=(\hat{c}_{KA,l},\hat{c}_{KB,l})^T$ and $\hat{c}_{\bar{K},l}=(\hat{c}_{\bar{K}B,l},\hat{c}_{\bar{K}A,l})^T$. In the numerical implementation, a Wilson term is introduced to eliminate fermion doubling by removing the Dirac cones at the edges of the Brillouin zone, namely, at $(k_x,k_y)=(\pm\pi/a,0)$, $(0,\pm\pi/a)$, and $(\pm\pi/a,\pm\pi/a)$, while retaining the Dirac cone at $(k_x,k_y)=(0,0)$ \cite{andp.202300081}, with the Wilson term given by $\hat{\mathcal{H}}_W=t_W\sum_l\{\hat{c}^\dagger_l[4-2\cos(k_ya)]\hat{\sigma}_z\hat{c}_l-\hat{c}^\dagger_l\hat{\sigma}_z \hat{c}_{l+1}+h.c.\}$. We choose $t_W=0.4t$ in our numerical calculations, and this choice does not affect the main conclusions of our work. To facilitate the subsequent derivation of the valley current, we express Eq.\ (\ref{eq:h1d}) in a general form as
\begin{align}
  \hat{\mathcal{H}}_{\mathrm{1D}}=\sum_l(\hat{c}^\dagger_l\hat{h}_{l}\hat{c}_l+\hat{c}^\dagger_{l+1}\hat{t}_{l+1,l}\hat{c}_{l}+h.c.),
\end{align}
where $\hat{h}_{l}$ denotes the on-site energy matrix at site $l$, while $\hat{t}_{l+1,l}$ denotes the coupling matrix between the ($l+1$)th and ($l$)th layers. Both matrices can be readily read off from Eq.\ (\ref{eq:h1d}).

The valley supercurrent across the junction can be obtained by summing over all $k_y$ modes, i.e., $J_{\mathrm{v}}(\phi)=\sum_{k_y}j_{\mathrm{v}}(\phi)$, 
where $j_{\mathrm{v}}(\phi)$ denotes the valley supercurrent for the 1D lattice model described by $\hat{\mathcal{H}}_{\mathrm{1D}}$. We first derive the current operator $\hat{j}_\tau$ ($\tau=K,\bar{K}$) from the continuity equation $\partial_t\hat{\rho}_{\tau,l}+\hat{j}_{\tau,l}-\hat{j}_{\tau,l-1}=0$, where $\hat{\rho}_{\tau,l}=e\hat{c}^\dagger_{\tau,l}\hat{c}_{\tau,l}$ is the charge density operator for $\tau$ valley at site $l$ \cite{Caroli_1971}. The time derivative of $\hat{\rho}_{\tau,l}$ can be obtained using the Heisenberg equation of motion, from which the current operator can be derived as follows
\begin{align}
    \hat{j}_{\tau,l}=\frac{ie}{\hbar}(\hat{c}^\dagger_{\tau,l}\hat{t}^{\dagger}_{\tau,l+1,l}\hat{c}_{\tau,l+1}-\hat{c}^\dagger_{\tau,l+1}\hat{t}_{\tau,l+1,l}\hat{c}_{\tau,l}),\label{eq:jv2}
\end{align}
where $\hat{t}_{\tau,l+1,l}$ denotes the $\tau$-diagonal component of the $\hat{t}_{l+1,l}$ matrix, i.e., $\hat{t}_{\tau,l+1,l}=\{\hat{t}_{l+1,l}\}_{\tau\tau}$. Thus, the valley current is given by 
\begin{align}
  j_{\mathrm{v}}&=\langle\hat{j}_{K,l}\rangle-\langle\hat{j}_{\bar{K},l}\rangle=\frac{ie}{\hbar}\sum_\tau \tau\langle \hat{c}^\dagger_{\tau,l}\hat{t}^{\dagger}_{\tau,l+1,l}\hat{c}_{\tau,l+1}-\hat{c}^\dagger_{\tau,l+1}\hat{t}_{\tau,l+1,l}\hat{c}_{\tau,l}\rangle,\label{eq:jv}
\end{align}
which can be further evaluated using the Matsubara Green's function \cite{PhysRevLett.98.107002}, defined as $\mathcal{G}_{\tau\tau',ij}(s-s')=-\langle \hat{T}_s\hat{c}_{\tau,i}(s)\hat{c}_{\tau',j}^\dagger(s')\rangle$. Here, $\tau,\tau^{\prime}$ and $i,j$ label the valley and lattice-site indices, respectively, $s,s^{\prime}$ denote imaginary time, and $\hat{T}_s$ is the time-ordering operator. We thus obtain the valley current in the Matsubara Green's function representation
\begin{align}
    j_{\mathrm{v}}&=\frac{ie}{\beta\hbar}\sum_{\omega_n}\mathrm{Tr}\big[\hat{t}^\dagger_{K,l+1,l}\mathcal{G}_{K,l+1,l}(i\omega_n)-\hat{t}_{K,l+1,l}\mathcal{G}_{K,l,l+1}(i\omega_n)-\hat{t}^\dagger_{\bar{K},l+1,l}\mathcal{G}_{\bar{K},l+1,l}(i\omega_n)+\hat{t}_{\bar{K},l+1,l}\mathcal{G}_{\bar{K},l,l+1}(i\omega_n)\big]\nonumber\\
    &=-\frac{ie}{\beta\hbar}\sum_{\omega_n}\mathrm{Tr}\big[\hat{\tau}_z \big(\mathcal{G}_{l,l+1}(i\omega_n)\hat{t}_{l+1,l}-\hat{t}^\dagger_{l+1,l} \mathcal{G}_{l+1,l}(i\omega_n)\big)\big],\label{eq:jv3}
\end{align} 
where $\beta=1/k_BT$ is the inverse temperature, $\omega_n=(2n+1)\pi/\beta$ is the Matsubara frequency, and $\mathcal{G}_\tau$ denotes the $\tau$-diagonal component of $\mathcal{G}$. In Eq.\ (\ref{eq:jv3}), the current is formulated in terms of the inter-cell Green's function, which can be further expressed in terms of on-cell quantities by means of Dyson's equation
\begin{align}
    \mathcal{G}_{l,l+1}(i\omega_n)=\mathcal{G}^{(L)}_{l,l}(i\omega_n)\hat{t}^\dagger_{l+1,l} \mathcal{G}_{l+1,l+1}(i\omega_n),\quad \mathcal{G}_{l+1,l}(i\omega_n)=\mathcal{G}_{l+1,l+1}(i\omega_n)\hat{t}_{l+1,l}\mathcal{G}^{(L)}_{l,l}(i\omega_n).\label{eq:dy}
\end{align}
Here $\mathcal{G}^{(L)}_{l_0,l_0}(i\omega_n)$ is the Matsubara Green's function of a system in which all lattice sites $l\geq l_0+1$ are deleted. Thus, Eq.\ (\ref{eq:jv3}) reduces to
\begin{align}
    j_{\mathrm{v}}&=\frac{e}{\beta\hbar}\sum_{\omega_n} 4\times\mathrm{Im}\mathrm{Tr}\{g_{K\bar{K}}\hat{t}^\dagger_{\bar{K}}\mathcal{G}_{\bar{K}K}\hat{t}_K\},
\end{align}
where we use the short notation $g=\mathcal{G}_{l,l}^{(L)}(i\omega_n)$, $\mathcal{G}=\mathcal{G}_{l+1,l+1}(i\omega_n)$, and $\hat{t}=\hat{t}_{l+1,l}$. Both $g$ and $\mathcal{G}$ can be calculated numerically using the recursive iteration method \cite{Sancho_1985,Sancho_1984}.

Furthermore, we can show more generally that the net \textit{charge} current in the system always vanishes at equilibrium. In contrast to the valley current, the charge current is given by the sum of the contributions from the two valleys:
\begin{align}
  j_{\mathrm{c}}&=\langle \hat{j}_{K,l}\rangle+\langle \hat{j}_{\bar{K},l}\rangle=\frac{ie}{\hbar}\sum_{\tau}\langle\hat{c}^\dagger_{\tau,l}\hat{t}^{\dagger}_{\tau,l+1,l}\hat{c}_{\tau,l+1}-\hat{c}^\dagger_{\tau,l+1}\hat{t}_{\tau,l+1,l}\hat{c}_{\tau,l}\rangle.\label{eq:jc}
\end{align}
Following the same procedure as that used to calculate the valley current, we obtain the charge current
\begin{align}
    j_{\mathrm{c}}&=-\frac{ie}{\beta\hbar}\sum_{\omega_n}\mathrm{Tr}\big[\hat{\tau}_0 \big(\mathcal{G}_{l,l+1}(i\omega_n)\hat{t}_{l+1,l}-\hat{t}_{l+1,l}^\dagger \mathcal{G}_{l+1,l}(i\omega_n)\big)\big]\nonumber\\
    &=-\frac{ie}{\beta\hbar}\sum_{\omega_n}\mathrm{Tr}\big[\hat{\tau}_0g(i\omega_n)\hat{t}^\dagger\mathcal{G}(i\omega_n)\hat{t}-\hat{\tau}_0\hat{t}^\dagger\mathcal{G}(i\omega_n)\hat{t}g(i\omega_n)\big]\nonumber\\
    &=-\frac{ie}{\beta\hbar}\sum_{\omega_n}\mathrm{Tr}\big\{[\hat{\tau}_0,g]\hat{t}^\dagger \mathcal{G}\hat{t}\big\}\nonumber\\
    &=0,\label{eq:jc1}
\end{align} 
where the second line follows from the Dyson's equation in Eq.\ (\ref{eq:dy}) with the site indices omitted, and $\hat{\tau}_0$ denotes the $4\times4$ identity matrix. Note that the derivation of Eq.\ (\ref{eq:jc1}) is general and does not depend on the specific forms of $\hat{t}$ and $\mathcal{G}$. This ensures that no spurious charge current signal arises in equilibrium, even in the presence of external fields that break time-reversal symmetry. Therefore, our model is more robust against external perturbations than previously reported dissipationless valley currents, including those induced by superconductivity \cite{PhysRevLett.121.226604,PhysRevB.103.125147,PhysRevB.110.024511} and those arising from tunneling in non-superconducting systems \cite{PhysRevLett.131.246301,fnmp-rtjx,PhysRevB.110.205406}, which are typically accompanied by a finite charge current.

\section{Reflection phase and bound state energy}
In this section, we present the detailed procedure for calculating the reflection phases and Andreev-like bound states using the continuum model.

\subsection{Scattering states in the continuum model}
The scattering states can be obtained by solving the secular equation $\hat{\mathcal{H}}\psi(x)e^{ik_yy}=E\psi(x)e^{ik_yy}$, where $k_y$ and $E$ are the conserved transverse wave number and the incident energy, respectively. In the Kekul\'e-Y distorted central region of the KGS junction, the scattering states for the $E_+$ band are given by ($\hbar v_F=1$)
\begin{align}
  \psi_{+}^{(>)}(x)=\begin{pmatrix}
    l_+\\k_0^++ik_y\\-k_0^++ik_y\\l_+
  \end{pmatrix}e^{ik_0^+x},\quad 
  \psi_{+}^{(<)}(x)=\begin{pmatrix}
    l_+\\-k_0^++ik_y\\k_0^++ik_y\\l_+
  \end{pmatrix}e^{-ik_0^+x},\label{eq:wk1}
\end{align}
where $k_0^+=\sqrt{l_+^2-k^2_y}$ is the longitudinal wave vector and $l_+=(E-U)/(1+\zeta)$. The superscripts ($>$) and ($<$) denote forward- and backward-propagating directions, respectively. For the $E_-$ band, the corresponding scattering states are given by
\begin{align}
  \psi_{-}^{(>)}(x)=\begin{pmatrix}
    -l_-\\-k_0^--ik_y\\-k_0^-+ik_y\\l_-
  \end{pmatrix}e^{ik_0^-x},\quad 
  \psi_{-}^{(<)}(x)=\begin{pmatrix}
    -l_-\\k_0^--ik_y\\k_0^-+ik_y\\l_-
  \end{pmatrix}e^{-ik_0^-x},\label{eq:wk2}
\end{align}
where $k_0^-=\sqrt{l_-^2-k_y^2}$ and $l_-=(E-U)/(1-\zeta)$. In the Kekul\'e-O distorted regions on either side of the junction, both of the conduction and valence bands are twofold degenerate, giving rise to two distinct scattering states for each propagation direction. The forward-propagating states are given by 
\begin{align}
  \psi_{1,j}^{(>)}(x)=\begin{pmatrix}
    Ee^{i\phi_j}\\(k_0+ik_y)e^{i\phi_j}\\\Delta\\0
  \end{pmatrix}e^{ik_0x},\quad 
  \psi_{2,j}^{(>)}(x)=\begin{pmatrix}
    (k_0-ik_y)e^{i\phi_j}\\Ee^{i\phi_j}\\0\\\Delta
  \end{pmatrix}e^{ik_0x},
\end{align}
where $k_0=\sqrt{E^2-\Delta^2-k_y^2}$ is the longitudinal wave vector and $j=L,R$. The backward-propagating states are given by
\begin{align}
  \psi_{1,j}^{(<)}(x)=\begin{pmatrix}
    Ee^{i\phi_j}\\(-k_0+ik_y)e^{i\phi_j}\\\Delta\\0
  \end{pmatrix}e^{-ik_0x},\quad 
  \psi_{2,j}^{(<)}(x)=\begin{pmatrix}
    -(k_0+ik_y)e^{i\phi_j}\\Ee^{i\phi_j}\\0\\\Delta
  \end{pmatrix}e^{-ik_0x}.\label{eq:wa2}
\end{align}

The backreflection phase acquired at each interface of the junction can be readily obtained by matching the propagating modes described above. Taking the right interface as an example, the wave function in the central Kekul\'e-Y region can be written as
\begin{align}
  \Psi_C(x)=\psi_{+}^{(>)}(x)+r\psi_{+}^{(<)}(x)+r^{\prime}\psi_{-}^{(<)}(x),\label{eq:pc}
\end{align}
where $r$ and $r^{\prime}$ denote the intraband and interband reflection amplitudes, respectively. In the Kekul\'e-O region, the wave function is given by
\begin{align}
  \Psi_R(x)=p_1\psi_{1,R}^{(>)}(x)+p_2\psi_{2,R}^{(>)}(x),
\end{align}
where $p_{1}$ and $p_2$ represent the corresponding transmission amplitude for $\psi_{1,R}^{(>)}$ and $\psi_{2,R}^{(>)}$. The undistorted graphene barrier of length $d$ at the interface is characterized by the dimensionless barrier strength $Vd/\hbar v_F$. We consider the thin-barrier limit, in which $V\to\infty$ and $d\to 0$, while keeping $Vd/\hbar v_F$ finite. At this stage, the wave functions on the two sides of the right boundary ($x=\ell$) are related by a transfer matrix $U_R$, as expressed by 
\begin{align}
  \Psi_C(\ell)=U_R\Psi_R(\ell).\label{eq:cc0}
\end{align}
The matrix $U_R$ encodes the information about the barrier and must also satisfy the current conservation condition. Once $U_R$ is specified (see the next subsection), solving Eq.\ (\ref{eq:cc0}) yields the intraband and interband reflection amplitudes, $r$ and $r'$, and their corresponding phases, $\arg(r)$ and $\arg(r')$.

\subsection{Boundary condition at the Kekul\'e-Y/O interface}
We next derive the transfer matrix $U_R$. Current conservation across the interface imposes the condition
\begin{align}
  \Psi_C^\dagger \hat{v}_{\text{Kekul\'e-Y}}\Psi_C=\Psi_R^\dagger \hat{v}_{\text{Kekul\'e-O}}\Psi_R,\label{eq:cc1}
\end{align}
where $\hat{v}_{\text{Kekul\'e-Y}}=\hbar v_F(\hat{\tau}_z\hat{\sigma}_x-\zeta\hat{\tau}_x\hat{\sigma}_z)$ and $\hat{v}_{\text{Kekul\'e-O}}=\hbar v_F\hat{\tau}_z\hat{\sigma}_x$ are the velocity operators along the junction for the Kekul\'e-Y and Kekul\'e-O regions, respectively. By introducing an auxiliary parameter $\lambda=\arctan(\zeta/i)$ and substituting Eq.\ (\ref{eq:cc0}) into Eq.\ (\ref{eq:cc1}), we obtain
\begin{align}
  U_R^\dagger[\hat{\tau}_z\hat{\sigma}_x\exp(-i\lambda\hat{\tau}_y\hat{\sigma}_y)]U_R=\cos\lambda\hat{\tau}_z\hat{\sigma}_x.\label{eq:cc2}
\end{align}
Setting $U_R=\sqrt{\cos\lambda}e^{-i\lambda\hat{\tau}_y\hat{\sigma}_y/2}\Xi_0$ allows us to factor out the exponential term in Eq.\ (\ref{eq:cc2}), yielding the following constraint
\begin{align}
  \Xi_0^\dagger\hat{\tau}_z\hat{\sigma}_x\Xi_0=\hat{\tau}_z\hat{\sigma}_x.\label{eq:mm0}
\end{align}
Assuming that $\Xi_0$ does not mix different valleys, we can further write 
\begin{align}
  \Xi_0=\begin{pmatrix}
    \mathcal{M}&\slashzero\\
    \slashzero&\mathcal{M}^\prime
  \end{pmatrix},\label{eq:mm1}
\end{align}
where $\slashzero$ denotes the $2\times2$ zero matrix, $\mathcal{M}$ and $\mathcal{M}'$ are the transfer matrices for the $K$ and $\bar{K}$ valleys, respectively. Substituting Eq.\ (\ref{eq:mm1}) into Eq.\ (\ref{eq:mm0}), we obtain the following relation that $\mathcal{M}$ and $\mathcal{M}'$ must satisfy:
\begin{align}
  \mathcal{M}^{(\prime)\dagger}\hat{\sigma}_x\mathcal{M}^{(\prime)}=\hat{\sigma}_x.\label{eq:mm2}
\end{align}
For brevity, we use $\mathcal{M}^{(\prime)}$ to denote either $\mathcal{M}$ or $\mathcal{M}'$ in Eq.\ (\ref{eq:mm2}). In addition, $\Xi_0$ should preserve both symplectic symmetry $\hat{\mathcal{S}}=i\hat{\tau}_z\hat{\sigma}_y\hat{\mathcal{K}}$ (acting as an effective time-reversal symmetry within a single valley) \cite{RevModPhys.80.1337} and time-reversal symmetry $\hat{\mathcal{T}}=\hat{\tau}_x\hat{\sigma}_x\hat{\mathcal{K}}$, which allow us to further determine the forms of $\mathcal{M}$ and $\mathcal{M}'$. The $\hat{\mathcal{S}}$ symmetry requires $\mathcal{M}^{(\prime)}$ to satisfy $\hat{\sigma}_y\mathcal{M}^{(\prime)*}\sigma_y=\mathcal{M}^{(\prime)}$, which allows $\mathcal{M}^{(\prime)}$ to be written as
\begin{align}
  \mathcal{M}^{(\prime)}=\gamma_0^{(\prime)}+\gamma_1^{(\prime)}\hat{\sigma}_x+\gamma_2^{(\prime)}\hat{\sigma}_y+\gamma_3^{(\prime)}\hat{\sigma}_z,\label{eq:mm3}
\end{align}
with $\gamma^{(\prime)}_{i}$ ($i=0,1,2,3$) being real parameters. By substituting Eq.\ (\ref{eq:mm3}) into Eq.\ (\ref{eq:mm2}), the parameters $\gamma_i^{(\prime)}$ can be determined
\begin{align}
  \gamma_0^{(\prime)}=\sqrt{1-(\gamma_1^{(\prime)})^2},\quad \gamma_2^{(\prime)}=\gamma_3^{(\prime)}=0,
\end{align}
which resulting in $\mathcal{M}=e^{i\beta\hat{\sigma}_x}$ and $\mathcal{M}^\prime=e^{i\beta^\prime\hat{\sigma}_x}$ with $\beta=\arcsin\gamma_1$ and $\beta^{\prime}=\arcsin\gamma_1^{\prime}$. The time-reversal symmetry requires $\beta$ and $\beta^{\prime}$ to satisfy $\beta^{\prime}=-\beta$, and thus $\Xi_0$ is finally obtained as
\begin{align}
  \Xi_0=\begin{pmatrix}
    e^{i\beta\hat{\sigma}_x}&\slashzero\\
    \slashzero&e^{-i\beta\hat{\sigma}_x}
  \end{pmatrix}.
\end{align}
Here the parameter $\beta$ corresponds to the barrier strength at the interface, with its value given by $\beta=Vd/\hbar v_F=V\delta_x/t$ \cite{PhysRevB.110.085301,PhysRevB.105.054502,PhysRevLett.97.217001}. This fully specifies the interface transfer matrix
\begin{align}
   U_R=\sqrt{\cos\lambda}e^{-i\lambda\hat{\tau}_y\hat{\sigma}_y/2}\begin{pmatrix}
    e^{i\beta\hat{\sigma}_x}&\slashzero\\
    \slashzero&e^{-i\beta\hat{\sigma}_x}
  \end{pmatrix}.
 \end{align}

\subsection{Andreev-like Bound state energy}
\begin{figure}[tp]
\begin{center}
\includegraphics[clip = true, width =0.95\linewidth]{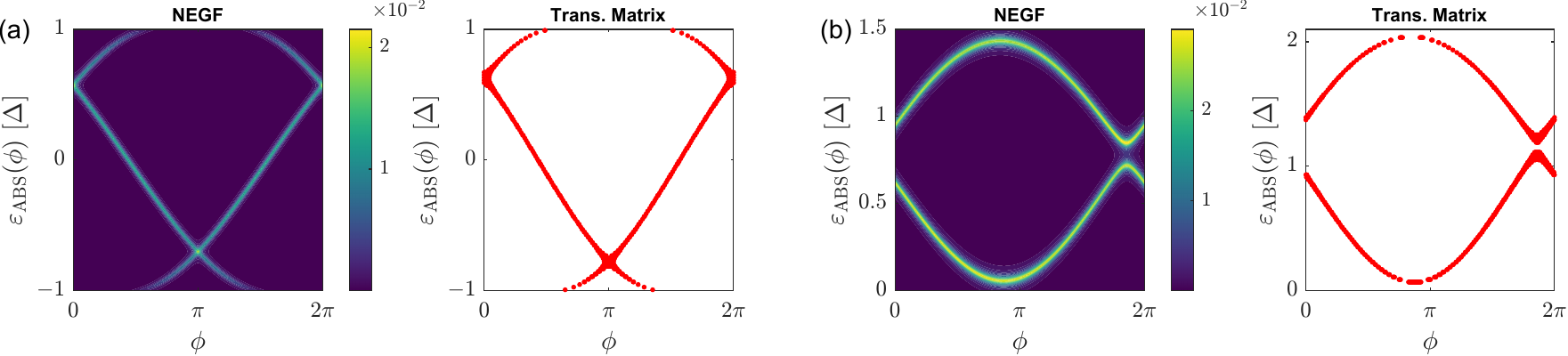}
\end{center}
\caption{$\varepsilon_{\mathrm{ABS}}$ as a function of $\phi$ at $\zeta=0.2$ for (a) $k_y=0$ and (b) $k_y=3$. In (a) and (b), the results on the left are calculated using the NEGF method, while those on the right are obtained using the transfer matrix method.}
\label{fig:s1}
\end{figure}

Similar to conventional Josephson junctions, multiple reflections within the central region of the KGS junction can give rise to Andreev-like bound states (ABS). The ABS energy ($\varepsilon_{\mathrm{ABS}}$) can be obtained by searching the peaks of the local density of states
\begin{align}
  \mathrm{LDOS}_l(E)=-\frac{1}{\pi}\operatorname{Im}\{\operatorname{Tr}\mathcal{G}^r_{l,l}(E)\},
\end{align}
where $\mathcal{G}^r$ is the retarded Green's function calculated by the standard nonequilibrium Green's function (NEGF) approach \cite{PhysRevB.57.10972,Sancho_1985,Sancho_1984}. Alternatively, ABS can be obtained using the transfer matrix approach within the scattering formalism \cite{RevModPhys.80.1337,PhysRevB.74.041401}, with the energy spectrum determined by the following equation
\begin{align}
  \det\begin{bmatrix}
   U^{-1}_RT_CU_L\psi_{1,L}^{(<)}(0),&U^{-1}_RT_CU_L\psi_{2,L}^{(<)}(0),& -\psi_{1,R}^{(>)}(\ell),&-\psi_{2,R}^{(>)}(\ell) 
  \end{bmatrix}=0.
\end{align}
Here $\det[\cdots]$ denotes the determinant of the $4\times 4$ matrix formed from the scattering states. $T_C$ denotes the transfer matrix of the central region, which can be directly constructed from Eqs.\ (\ref{eq:wk1}) and (\ref{eq:wk2}) \cite{PhysRevB.74.041401,PhysRevB.86.134522}, while $U_L$ is the transfer matrix at the left boundary, obtained using the same procedure as for $U_R$. Figure\ \ref{fig:s1} presents the ABS energy spectra at $\zeta=0.2$ obtained using the two approaches. At $k_y=0$, the NEGF and transfer matrix methods yield identical results, as shown in Figure\ \ref{fig:s1}(a). At $k_y=3$, the two methods also produce the same energy spectrum, although the NEGF spectrum exhibits an overall energy offset of approximately $0.5$ relative to the transfer matrix result. This shift originates from the regularization introduced by the Wilson mass at finite $k_y$ and does not affect the resulting current.

\section{Derivation of Eq.\ (10) in the main text}

From Eqs.\ (6-8) in the main text, we obtain the valley supercurrent at a given $k_y$  
\begin{align}
  j_{\mathrm{v}}(\phi)&=j_0\sum_s\frac{t_s\sin2\theta_s}{\sqrt{1-t_s\sin^2\theta_s}}\big(1-sw\cos(\phi/2)\big)\nonumber\\
  &=j_0\sum_sF_{t_s}\big(x-sw\sin(x-q)\big)\times\big(1-sw\cos(x-q)\big)\nonumber\\
  &=j_0\sum_s\{F_{t_s}(x)-sw[\sin(x-q)F'_{t_s}(x)+\cos(x-q)F_{t_s}(x)]\}+\mathcal{O}(\mathcal{C}^2),\label{eq:isv}
\end{align}
where $F_t(x)=\frac{t\sin 2x}{\sqrt{1-t\sin^2x}}$, $x=\phi/2+\delta k_F\ell$, $j_0=e\Delta/2\hbar$, and $q=\delta k_F\ell$. By defining 
\begin{align}
  A(x)=F_{t_+}(x)+F_{t_-}(x),\quad B(x)=F_{t_+}(x)-F_{t_-}(x),
\end{align}
Eq.\ (\ref{eq:isv}) can be further simplified to
\begin{align}
  j_{\mathrm{v}}=j_0\times[A(x)+w a(x)]+\mathcal{O}(w^2),
\end{align}
with $a(x)=-[\sin(x-q)B'(x)+\cos(x-q)B(x)]$. Note that both $A(x)$ and $B(x)$ are odd functions. In the zeroth-order approximation ($w=0$), let the forward current attain its maximum at $x=x_0$, such that $A(x_0)=A_m$. By odd symmetry, the reverse current reaches its maximum magnitude at $x=-x_0$, with $A(-x_0)=-A_m$. Consequently, the forward and reverse critical valley currents to first order are then given by
\begin{align}
  j_{\mathrm{v}}^{c+}=j_0\times[A_m+wa(x_0)],\quad |j_{\mathrm{v}}^{c-}|=j_0\times[A_m-wa(-x_0)],
\end{align}
resulting in the difference between the forward and reverse critical valley currents
\begin{align}
  \delta j_{\mathrm{v}}^c&=j_{\mathrm{v}}^{c+}-|j_{\mathrm{v}}^{c-}|\nonumber\\
  &=2j_0w[\cos x_0\sin qB'(x_0)-\sin x_0\sin qB(x_0)]+\mathcal{O}(w^2)\nonumber\\
  &=w\sin(\delta k_F\ell)\times Q+\mathcal{O}(w^2),
\end{align}
where $Q=(e\Delta/\hbar)[B'(x_0)\cos x_0-B(x_0)\sin x_0]$.